\def\junk#1{}
\documentclass[a4paper, 10pt, conference]{ieeeconf}      

\IEEEoverridecommandlockouts                              

\usepackage{color,soul,framed}
\usepackage{setspace} 
\usepackage[usenames,dvipsnames,svgnames,table]{xcolor}
\usepackage{graphics} 
\usepackage{epsfig} 
\usepackage[caption=false,font=footnotesize]{subfig}
\usepackage{times} 
\usepackage[cmex10]{amsmath}
\usepackage{amssymb}
\usepackage{amstext}
\usepackage{url}
\usepackage{algorithm}
\usepackage{algorithmic}
\usepackage{mdwmath}
\usepackage{mdwtab}

\usepackage{epstopdf}
\usepackage{multirow}

\usepackage[switch]{lineno}

\author{
\begin{tabular}{cccc}
\vspace{-0.05in}
Flavio Esposito\thanks{Flavio Esposito's work was done while at Boston University.} &&   Ibrahim Matta   \\
{\small \; fesposito@exegy.com } && {\small \; matta@cs.bu.edu } \\
{\normalsize Exegy Inc. } && \normalsize{Computer Science Department} \\
{\normalsize St. Louis, MO } && \normalsize{Boston University, MA}  
\end{tabular}
}

\title{\Large \bf
A Decomposition-based Architecture for Distributed Virtual Network Embedding 
}

\begin{document}


\maketitle
\thispagestyle{empty}
\pagestyle{empty}

\begin{abstract}
Network protocols have historically been developed on an ad-hoc basis, and cloud computing is no exception. A fundamental management protocol, not yet standardized, that cloud providers need to run to support wide-area virtual network services is the virtual network (VN) embedding protocol. 

In this paper, we use decomposition theory to provide a unifying architecture  for the VN embedding problem. We show how our architecture subsumes existing solutions, and how it can be used by cloud providers to design a distributed VN embedding protocol that adapts to different scenarios, by merely instantiating different decomposition policies. We analyze key representative tradeoffs via simulation, and with our VN embedding testbed that uses a Linux system architecture to reserve virtual node and link capacities. 
In contrast with existing VN embedding solutions, we found that partitioning a VN request not only increases the signaling overhead, but may decrease cloud providers' revenue.
\end{abstract}
\vspace{-2mm}

\section{Introduction}\label{sec:intro}

The cloud computing market is rapidly becoming dominated by a small set of public infrastructure providers, that profit from concurrently running multiple customized (virtual network) services on their shared or leased infrastructure. 
One of the fundamental management protocols, not yet standardized, that cloud providers need to run to support virtual network services is the virtual network (VN) embedding protocol~\footnote{We call service providers the players that do not own the infrastructure but provide a (cloud-based) service. Infrastructure providers own instead the physical network resources. A cloud provider can be a lessor or a lessee of the network infrastructure, and can act as both service and infrastructure provider.}. 
 Running such protocol requires solving the NP-hard problem of matching constrained virtual networks on the physical network (overlay) of the infrastructure provider. The virtual network embedding problem consists of three interacting mechanisms: $(i)$ resource discovery, where the space of the available potentially hosting physical (or overlay) resources is sampled or exhaustively searched; $(ii)$ virtual network mapping, where a subset of  available physical resources is chosen as a candidate to potentially host the requested virtual network, and $(iii)$ allocation, where each virtual node is bound to a physical node, and each virtual link to at least one loop-free physical path. 

Distributed embedding solutions are useful to single cloud providers to enable virtual network services that span a wide geographical area (see $e.g.$ the GENI testbed~\cite{GENI}), but many envision also a {\it ``cloud marketplace"} where vendors of software, hardware, and services collectively participate in operating open-cloud solutions~\cite{azer-ocx}. 
Distributed solutions that allow service and infrastructure providers to collectively embed a VN already exist~\cite{Houidi-distributedVNM,V-Mart,Polyvine}. For example, some solutions focus on the desirable property of letting infrastructure providers use their own (embedding) policies~\cite{Polyvine}, while others rely on truthfulness of virtual resource auctions~\cite{V-Mart}.  
Although they have systematic logic behind their design, such distributed solutions are restricted to a subset of the three virtual network embedding tasks, they have performance ($e.g.$ convergence speed or resource utilization) tightly determined by the chosen heuristic, and they are limited to a single distribution model --- the type and amount of information propagated to embed a VN. 
This is because their heuristic design is tailored to specific provider goals, $e.g.$, minimize the virtual node or path migrations, or maximize the number of VNs to be hosted. Moreover, such heuristics are also tailored to specific allocation models $e.g.$, best effort or dictated by a Service Level Agreement (SLA). 
In summary, a VN embedding solution valid for all providers' goals, that tackles all possible Service Level Objectives (SLOs) --- the technical requirements within an SLA --- probably cannot exist. 
Instead of designing new protocols in response to new sets of provider's goals, services, or SLAs, the VN embedding problem may be holistically analyzed and systematically solved as a distributed solution to some global optimization, in the form of a general network utility maximization problem.
To this aim, in this paper we present the following contributions:

\noindent
{\bf Architecture and decompositions.}  We analyze the design challenges in embedding a  VN (Section~\ref{sec:model}), and we use decomposition theory~\cite{boyd-book} to construct an analytic foundation for the tradeoff analysis of distributed embedding solutions. 
Our architecture can be used to subsume existing solutions, and to design novel VN embedding protocols by selecting the appropriate variables to fix, when using a primal decomposition, and the appropriate constraints to relax, when using a dual decomposition. 
 For example, primal decompositions can be used to model different VN partitioning policies,~\footnote{The VN partitioning problem (or VN graph splitting) is the (NP-hard~\cite{Houidi2011}) problem of splitting a VN into multiple connected subsets of virtual nodes and links.} and  dual decomposition techniques to relax different constraints isolating the three VN embedding mechanisms into different architectural blocks (Section~\ref{sec:primalVsDual}). 
Fixing the decision variables associated with the virtual links is equivalent to optimize the embedding of virtual nodes first, and the virtual link embedding later as in~\cite{pathsplitting}.
As another example, by optimizing the embedding of a given VN partition first, $e.g.$, the partition with the highest virtual node and link capacities, and then the next partition, our architecture subsume the heuristic used in~\cite{Houidi-distributedVNM}. 
Finally, solving the VN embedding by applying primal and later dual decomposition is equivalent to the approach used in~\cite{CAD}: a distributed VN embedding protocol first partitions the VN, and then the prices on virtual resources are exchanged among physical nodes so that congested physical nodes pay a higher price to host a virtual node of the partition.

\noindent
 {\bf Comparing different embedding policies.} Using simulations and a Linux-based VN embedding testbed~\cite{myPhDThesisTR}, we evaluate few key design tradeoffs for a set of representative policy instantiations of our architecture.
Tradeoffs include message passing overhead against convergence speed of the iterative methods used to solve the VN embedding problem (modeled in Section~\ref{sec:model}), and different VN partitioning policies  (Section~\ref{sec:primalVsDual}). 
Each infrastructure provider process of our testbed includes the modules of a prototype implementation of each of the three VN embedding mechanisms.
Each emulated virtual node is a user-level process that has its own virtual Ethernet interface(s), created and installed with {\tt ip link add/set}, and it is attached to an Open vSwitch~\cite{openvswitch} running in kernel mode to switch packets across virtual interfaces (Section~\ref{sec:eval}).

\vspace{-1mm}
\section{Model and Design Challenges}\label{sec:model}


In this section we describe the VN embedding problem as a general network utility maximization problem. Previous models have used optimization theory to capture different objectives and constraints of the VN embedding problem (see $e.g.$~\cite{ChowdhuryTON,Houidi2011}). Our model captures all three mechanisms of the VN embedding problem: resource discovery,  virtual network mapping,  and allocation.  We begin the section defining such mechanisms and describing some of the challenges associated with designing a distributed VN embedding solution.  

 \emph{Resource discovery} is the process of monitoring the state of the substrate (physical or overlay) resources using sensors and other measurement processes. The monitored states include processor loads, memory usage, network performance data, etc. 
The major challenge in designing a resource discovery system is presented by the 
different VN's arrival rates and durations that the cloud provider might need to support: the lifetime of a VN can range from a few seconds (in the case of cluster-on-demand services) to several months (in the case of a VN hosting a GENI~\cite{GENI} experiment looking for new adopters to opt-in). 
In wide-area testbed applications, VNs are provided in a best-effort manner, and the inter-arrival time between VN requests and the lifetime of a VN are typically much longer than the embedding time, so designers may assume complete knowledge of the network state, and ignore the overhead of resource discovery and the VN embedding time. On the other hand, in applications with higher churns, $e.g.$, cluster-on-demand such as financial modeling, anomaly analysis, or heavy image processing, where Service Level Agreements (SLAs) require short response time, it is desirable to reduce the VN embedding time, and employ limited resource discovery to reduce overhead.

{\it Virtual network mapping} is the step that matches VN requests with the available resources, and selects some subset of the resources that can potentially host the virtual network. 
Due to the combination of node and link constraints, this is the most complex of the virtual network embedding tasks. In fact the problem is NP-hard~\cite{mappingNPhard}. These constraints include intra-node constraints ($e.g.$, desired physical location, processor speed, storage capacity, type of network connectivity), as well as inter-node constraints ($e.g.$, VN topology).

Designing a VN mapping algorithm is also challenging. Within a small enterprise physical network for example, embedding virtual nodes and virtual links separately may be preferable to adapt to the physical network load with minimal virtual machine or path migrations~\cite{pathsplitting}. If the goal is instead to increase physical network utilization, virtual network mapping solutions that simultaneously embed virtual nodes and links may be preferable~\cite{ChowdhuryTON,isomorphism-visa09}. The heuristic used to partition the VN, that is the  input of the VN mapping algorithm, changes the space of solutions, the embedding time, or both~\cite{myPhDThesisTR}.

{\it Allocation} involves assigning (binding) one set of all physical (or overlay) resources among all those that match the VN query, to the VN.  If the resource allocation step fails, the matching step should be reinvoked. 
The allocation step can be a single shot process, or it can be repeated periodically to either assign or reassign different VN partitions, acquiring additional resources on a partial VN that has already been embedded (allocated).

The design challenges of the VN embedding are both architectural, $i.e.$, who should make the binding decisions, and algorithmic, $i.e.$, how should the binding occur.
A centralized third party provider can be in charge of orchestrating the binding process collecting information by (a subset of) multiple infrastructure providers~\cite{Houidi2011,cabernet,Polyvine,pathsplitting}, or the decision can be fully distributed~\cite{CAD,Houidi-distributedVNM,V-Mart}, using a broker~\cite{SHARP}, an auction mechanism~\cite{V-Mart}, First Come First Serve~\cite{sword-journal}, or maximizing some notion of utility, of a single service provider~\cite{pathsplitting} or of a set of infrastructure providers~\cite{CAD}. 
In summary, the design space of a VN embedding solution is large and unexplored, and many interesting solutions and tradeoff decisions are involved in this critical cloud resource allocation problem.

The design challenges are exacerbated by the interaction among the three mechanisms (phases).
The VN embedding problem is in fact a closed feedback system, where the three tasks are solved repeatedly; the solution at any given iteration affects the space of feasible solutions in the next iteration: the resource discoverer(s) return(s) a subset of the available resources to VN mapper(s). Subsequently, a list of candidate mappings are passed to the allocator(s), that decide(s) which physical (or overlay) resources are going to be assigned to each VN. After a successful binding, the allocator processes communicate with the resource discovery processes, so that future discovery operations are aware of the unavailable resources.

\noindent
{\bf Modeling Virtual Network Embedding.}
We model the VN embedding problem with a network utility maximization problem. In particular, we assume that {\it Pareto optimality} is sought among physical nodes, possibly belonging to different infrastructure or cloud providers. We maximize 
$\sum_i U_i$, where $U_i$ is a general utility function, measured on each hosting node $i$. Such function could depend on one, or all the VN embedding phases. 
In our model we assume that a VN request $j$ contains $\gamma_j$ virtual nodes, and $\psi_j$ virtual links;  hence, in order to embed $j$, the discovery system needs  to find at least $\gamma_j$ hosting nodes (constraint~\ref{1a}) and $\psi_j$ virtual links (constraint~\ref{1b}) to give at least one candidate to the VN mapper(s).
We model the result of the discovery mechanism with $n_{ij}^{P}$ and $p_{kj}$, equal to one if the hosting node $i$, and physical loop-free path $k$, respectively, were available, and zero otherwise. An element is available if a discovery operation is able to find it, given a set of protocol parameters, $e.g.$, find all loop-free paths within a given deadline, or find as many available physical nodes as possible within a given number of hops.  Similarly, we model the VN mapping mechanism with other two binary variables, $n_{ij}^{V}$  and $l_{kj}$, equal to one if a virtual instance of physical node $i$ and physical loop-free path $k$, respectively, are assigned to the VN request $j$, and zero otherwise.
Constraints ($1a$) and ($1b$) refer to the discovery, constraints ($1c$) and ($1d$) refer to the VN mapping, while ($1g$) and ($1h$) are the standard {\it set packing problem} constraints, and refer to the allocation, given a physical node capacity $C^{n}_i$, and the capacity of each loop-free physical path $C^{l}_k$. The VN embedding can be hence modeled as follows: 
\vspace{-2mm}
\begin{subequations}\label{sliceembedding}
\begin{eqnarray} \nonumber
{\rm maximize} & \displaystyle\sum_{i = 1}^{N_p} U_i (n_{ij}^P,p_{kj}, n_{ij}^V,l_{kj} , y_j)  \\  \label{obj}
{\rm  subject \; \rm to} & \displaystyle\sum_{i \in N} n_{ij}^P \geq  \gamma_j \;\; \forall j  \label{1a} \\
&				 \sum_{k \in \mathcal{P}} p_{kj} \geq  \psi_j \;\; \forall j  \label{1b} \\ 
&				\displaystyle\sum_{i \in N} n_{ij}^V =\gamma_j  \;\; \forall j  \\
&				\displaystyle\sum_{k \in \mathcal{P}} l_{kj} = \psi_j \;\; \forall j \\ 
&				 n_{ij}^V \leq n_{ij}^P  \;\;  \forall i  \;\;  \forall j \\ 
&				 l_{kj} \leq p_{kj}  \;\;  \forall k \;\;  \forall j  \\ 
& 				\displaystyle\sum_{j \in J} n_{ij}^V y_j \leq C_i^n \;\;  \forall i \\
&				   \displaystyle\sum_{j \in J} l_{kj} y_j \leq C_k^l  \;\; \forall k \\ 
&				  y_j \leq n_{ij}^V  \;\; \forall i \;\;  \forall j\\
&				  y_j \leq l_{kj}  \;\; \forall k  \;\; \forall j \\
&				 y_j, n_{ij}^P, p_{kj}, n_{ij}^V, l_{kj},  \in \{0,1\}  \;\;  \forall \;\; i , j , k
\end{eqnarray}
\end{subequations}
Constraints ($1e - 1f$) and ($1i- 1j$) are called {\it complicating constraints}, as they complicate the problem binding the three mechanisms together;  without those constraints, each of the VN embedding mechanism could be solved separately from the others,  $e.g.$, by a different architecture component. 

The existential constraints $(1k)$ could be relaxed in the interval $[0,1]$; in this case, the discovery variables could represent the fraction of available resources, while the mapping and allocation variables could model partial assignments; it is in fact feasible to virtualize a node on multiple servers, and a link on multiple loop-free physical paths~\cite{pathsplitting}.

\section{Virtual Network Embedding \\ Decomposition Architecture}\label{sec:dec}

Due to the rich structure of problem~(\ref{sliceembedding}), many different decompositions are possible.  
 Each alternative decomposition leads to a different distributed algorithm, with potentially different desirable properties. 
%
 The choice of the adequate decomposition method and distributed algorithm for a particular problem depends on the infrastructure providers' goals, and on the offered service or application. 
The idea of decomposing problem~(\ref{sliceembedding}) is to convert it into equivalent formulations, where a master problem interacts with a set of subproblems. Decomposition techniques can be classified into {\it primal} and {\it dual}. Primal decompositions are based on decomposing the original primal problem~(\ref{sliceembedding}), while dual decomposition methods are based on decomposing its dual. 
In a primal decomposition, the master problem allocates the existing resources by directly assigning to each subproblem the amount of resources that it can use. Dual decomposition methods instead correspond to a resource allocation via pricing, $i.e.$, the master problem sets the resource price for all the subproblems, that independently decide if they should host the virtual resources or not, based on such prices.

{\it Primal decompositions} are applicable to problem~(\ref{sliceembedding}) by an iterative {\it partitioning} of the decision variables into multiple subsets. Each partition set is optimized separately, while the remaining variables are fixed.
For example, we could first optimize the set of virtual node variables $n_{ij}^V$ in a node embedding phase, fixing the virtual link variables $l_{kj}$, and then optimize the virtual links in a path embedding phase, given the optimal value of the variables $n_{ij}^{V}$, obtained from the node embedding, as done in~\cite{CAD,pathsplitting}.
Alternatively, a distributed VN embedding algorithm could simultaneously optimize both virtual node and virtual link embedding for subsequent VN partitions, $e.g.$, sorting first the partitions by the highest requested virtual node and virtual link capacity, as in~\cite{Houidi-distributedVNM}.
Primal decompositions can also be applied with respect to the three VN embedding mechanisms. For example, by fixing the allocation variables, the embedding problem can be solved by optimizing the discovery and VN mapping first as in~\cite{Polyvine}, or by optimizing the discovery variables ${n}_{ij}^P$ and ${ p}_{kj}$  first, and then simultaneously the mapping and allocation variables later as in~\cite{sword}.

\begin{figure}[t!]
\centering
\includegraphics[width=.82\columnwidth]{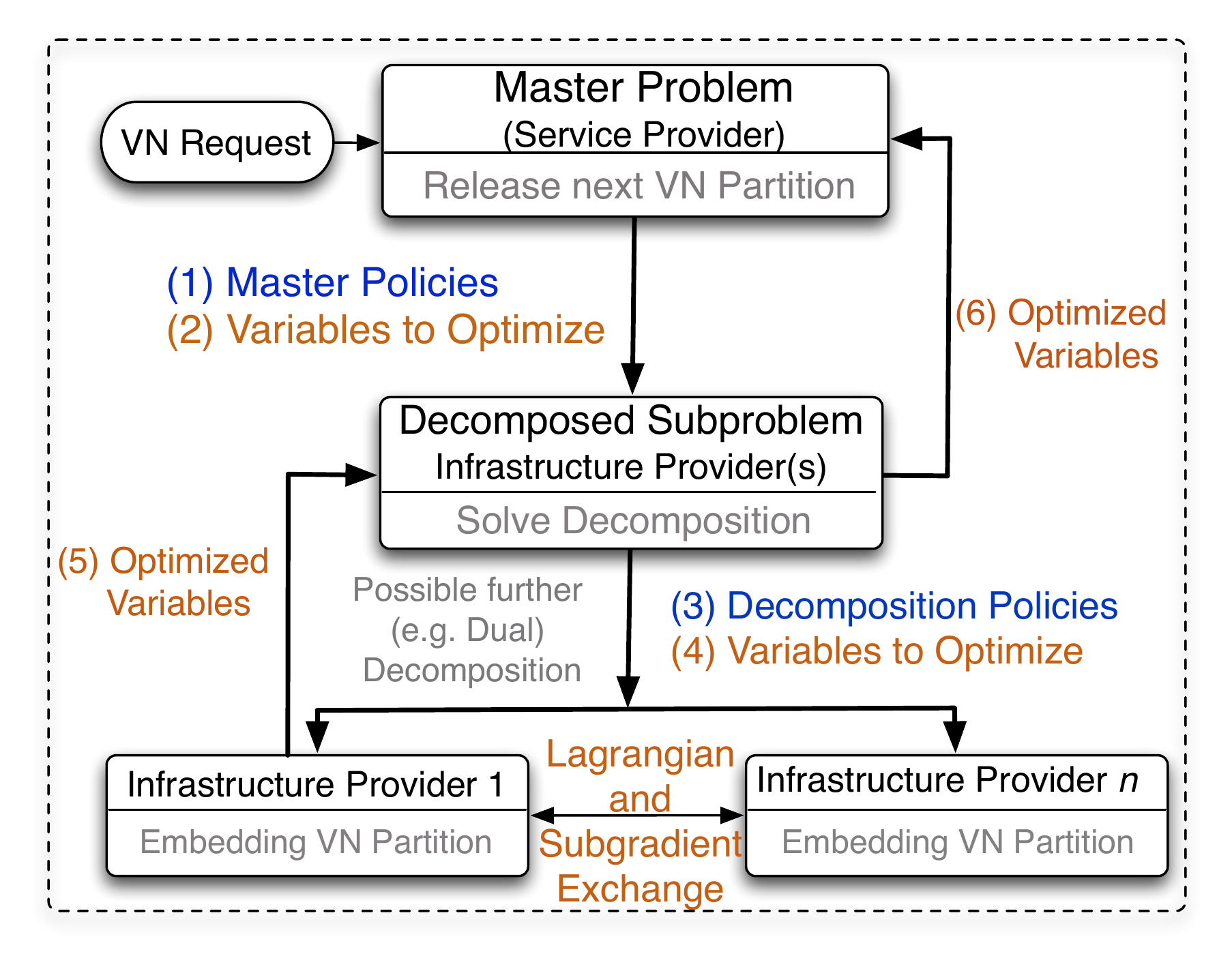}
\caption{Decomposition-based virtual network embedding architecture: different embedding solutions can be modeled via primal and dual decompositions. The service provider instantiates a problem formulation according to its policies (1),
and picks an objective function $U$ (2). The infrastructure provider processes solve the decomposed subproblems, possibly further decomposing them (3-4). Finally, the optimal embedding variables are returned to the service provider (5-6), that eventually releases the next VN.}
\label{fig:architecture}
\vspace{-5mm}
\end{figure}

{\it Dual decomposition} approaches are based on decomposing the Lagrangian function formed by augmenting the master problem with the relaxed constraints.
Even in this case, it is possible to obtain different decompositions by relaxing different sets of constraints, hence obtaining different distributed VN embedding algorithms.  For example, by relaxing constraints  
($\ref{sliceembedding}$i) and ($\ref{sliceembedding}$j), we can model solutions that separate the VN mapping and allocation phases, such as~\cite{SHARP,bellagio}.
Regardless of the number of constraints that are relaxed,   dual decompositions are different than primal in the amount of required parallel computation (all the subproblems could be solved in parallel), and the amount of message passing between one phase and the other of the iterative method. The dual master problem communicates to each subproblem the shadow prices, $i.e.$, the Lagrangian multipliers, then each of the subproblems (sequentially or in parallel) is solved, and  the optimal value is returned, together with the subgradients. 
It is also possible to devise VN embedding solutions in which both primal and dual decompositions are used. 

In general, a service provider can instantiate a set of policies at the {\it master problem}, after receiving a VN embedding request,  dictating the order in which the variables need to be optimized and on which VN partition. The subproblems resulting from the decomposition can also instantiate other sets of decomposition policies, to decide which variables are to be optimized next, in which order, or even further decomposing the subproblems (Figure~\ref{fig:architecture}).

\section{Primal versus Dual Decompositions}\label{sec:primalVsDual}

In this section we analyze the tradeoffs between primal and dual decompositions, for a simple VN embedding subproblem. We later use this case study to show the results of a tradeoff analysis between optimality and speed of convergence of the iterative method used by a CPLEX solver.
We consider a subproblem of problem~(\ref{sliceembedding}): the virtual node embedding problem, 
where the VN request is split in two partitions. The problem can be formulated as follows:
\vspace{-2mm}
\begin{subequations}\label{new-nodeEmbedding}
\begin{eqnarray} \nonumber
{\rm \displaystyle\max_{ u,  v} } &  c^{T} u + \tilde{c}^{T} v & \\  
{\rm  subject \; \rm to} & A u  \leq b &   \label{nonComplicatingFirst}\\
&		             \tilde{A} v   \leq \tilde{b} &  \label{nonComplicatingSecond}  \\
&				 F u + \tilde{F} v \leq h  \label{complicating}
\end{eqnarray}
\end{subequations} 
where $u$ and $v$ are the sets of decision variables referring to the first and to the second VN partition, respectively;  $F$ and $\tilde{F}$ are the matrices of capacity values for the virtual nodes in the two partitions, and $h$ is the vector of all physical node capacity limits.
The constraints~(\ref{nonComplicatingFirst}) and~(\ref{nonComplicatingSecond}) capture the separable nature of the problem into the two partitions. Constraint~(\ref{complicating}) captures the complicating constraint. 
\vspace{1mm}

\noindent
{\bf Embedding by Primal Decomposition.} By applying primal decomposition to problem~(\ref{new-nodeEmbedding}), we can separately solve two subproblems, one for each VN partition, by introducing an auxiliary variable $z$, that represents the percentage of physical and virtual resource allocated to each subproblem. The original problem~(\ref{new-nodeEmbedding}) is equivalent to the following master problem:
\vspace{-2mm}
 \begin{equation}
{\rm \displaystyle\max_{ z} } \ \phi(z) + \tilde{\phi}(z)
\label{masterPrimal}
  \end{equation}
\vspace{-2mm}
where:
\vspace{-2mm}
\begin{subequations}\label{slave-p1}
\begin{eqnarray} 
{\rm \displaystyle\phi(z)} = & \sup_{u}  c^{T} u \\
{\rm  subject \; \rm to} & A u  \leq b &  \\
&				 F u  \leq z  
\end{eqnarray}
\end{subequations} 
\vspace{-2mm}
and
\vspace{-2mm}
\begin{subequations}\label{slave-p2}
\begin{eqnarray} 
{\rm \displaystyle\tilde{\phi}(z)} = & \sup_{v}  \tilde{c}^{T} v \\
{\rm  subject \; \rm to} & \tilde{A} v  \leq \tilde{b} &   \\
&				 \tilde{F} v  \leq h - z.  
\end{eqnarray}
\end{subequations} 
The primal master problem~(\ref{masterPrimal}) maximizes the sum of the optimal values of the two subproblems, over the auxiliary variable $z$. After $z$ is fixed, the subproblems (\ref{slave-p1}) and (\ref{slave-p2}) are solved separately, sequentially or in parallel, depending on the cloud provider's policy. The master algorithm updates $z$, and collects the two subgradients, independently  computed by the two subproblems. 
%
To find the optimal $z$, we use a subgradient method.
%
In particular, to evaluate a subgradient of $\phi(z)$ and $\tilde{\phi}(z)$, we first find the optimal dual variables $\lambda^{\star}$ for the first subproblem subject to the constraint  $F u  \leq z$. Simultaneously (or sequentially), we find the optimal dual variables $\tilde{\lambda}^{\star}$ for the second subproblem, subject to the constraint  $\tilde{F} v  \leq h - z$. The subgradient of the original master problem is therefore $g = - \lambda^{\star}(z) + \tilde{\lambda}^{\star}(z)$; that is,  $g \in \partial (\phi(z) + \tilde{\phi}(z))$.\footnote{For the proof 
please refer to $\S 5.6$ of~\cite{boyd-book}.}
The primal decomposition algorithm, combined with the subgradient method for the master problem is repeated, using a diminishing step size, until a stopping criteria is reached (Procedure~\ref{alg:primal}).
\floatname{algorithm}{Procedure}
\begin{algorithm}[t]                   
\caption{\small{Distributed Embedding by Primal Decomposition }}
\begin{algorithmic}[1]              
\STATE { Given $z_t$ at iteration $t$, solve subproblems to obtain optimal embedding $\phi$ and $\tilde{\phi}$ for each VN partition, and dual variables \; $\lambda^{\star}(z_t)$ and $\tilde{\lambda}^{\star}(z_t)$ }
\STATE {Send/Receive $\phi$, $\tilde{\phi}$, $\lambda^{\star}$ and $\tilde{\lambda}^{\star}$ }
\STATE{Master computes subgradient $g(z_t) = -\lambda^{\star}(z_t)+ \tilde{\lambda}^{\star}(z_t)$ }
\STATE{Master updates resource vector  $z_{t+1} = z_t - \alpha_t g$}
\end{algorithmic}
\label{alg:primal}
\end{algorithm} 
The optimal Lagrangian multiplier associated with the capacity of the physical node $i$,  $- \lambda^{\star}_i$, tells us how much worse the objective of the first subproblem would be, for a small (marginal) decrease in the capacity of the physical node $i$. $\tilde{\lambda}^{\star}_i$ tells us how much better the objective of the second subproblem would be, for a small (marginal) increase in the capacity of physical node $i$. Therefore, the primal subgradient  $g(z) = -\lambda(z)+ \tilde{\lambda}(z)$ tells us how much better the total objective would be if we transfer some physical capacity of physical node $i$ from one subsystem to the other.  At each step of the subgradient method, more capacity of each physical node is allocated to the subproblem with the larger Lagrange multiplier. This is done with an update of the auxiliary variable $z$. The resource update $z_{t+1} = z_t - \alpha_t g$ can be interpreted as shifts of some of the capacity to the subsystem that can better use it for the global utility maximization.

\vspace{1mm}
\noindent
{\bf Embedding by Dual Decomposition.}
An alternative method to solve problem~(\ref{new-nodeEmbedding}) is to use dual decomposition, relaxing the coupling capacity constraint~(\ref{complicating}). From problem~(\ref{new-nodeEmbedding}) we form the partial Lagrangian function:

\vspace{-5mm}
\begin{subequations}
\begin{eqnarray} 
L(u,v,\lambda) = c^T u + \tilde{c}^T v + \lambda^T (Fu + \tilde{F}v - h) 
\label{lagrangian}
\end{eqnarray}
\end{subequations} 
\vspace{-2mm}
Hence, the dual function is:
%
\begin{subequations}
\begin{eqnarray} 
q(\lambda) = \inf_{u,v} \{ L(u,v,\lambda) | Au \leq b, \tilde{A} v \leq \tilde{b} \} \\
 = -\lambda^T h + \inf_{Au \leq b} (F^T \lambda + c)^T u + \inf_{\tilde{A} v \leq \tilde{b}} (\tilde{F}^T\lambda + \tilde{c})^T v \nonumber,
\label{dualFunction}
\end{eqnarray}
\end{subequations} 
\vspace{-2mm}
and the dual problem is:
\begin{subequations}\label{dual}
\begin{eqnarray} 
\max_{\lambda} & q(\lambda) \\
{\rm  subject \; \rm to} & \lambda \geq 0 \nonumber,
\end{eqnarray}
\end{subequations}

\begin{figure*}[t!]
\centering
\subfloat[]{\includegraphics[width=.51\columnwidth]{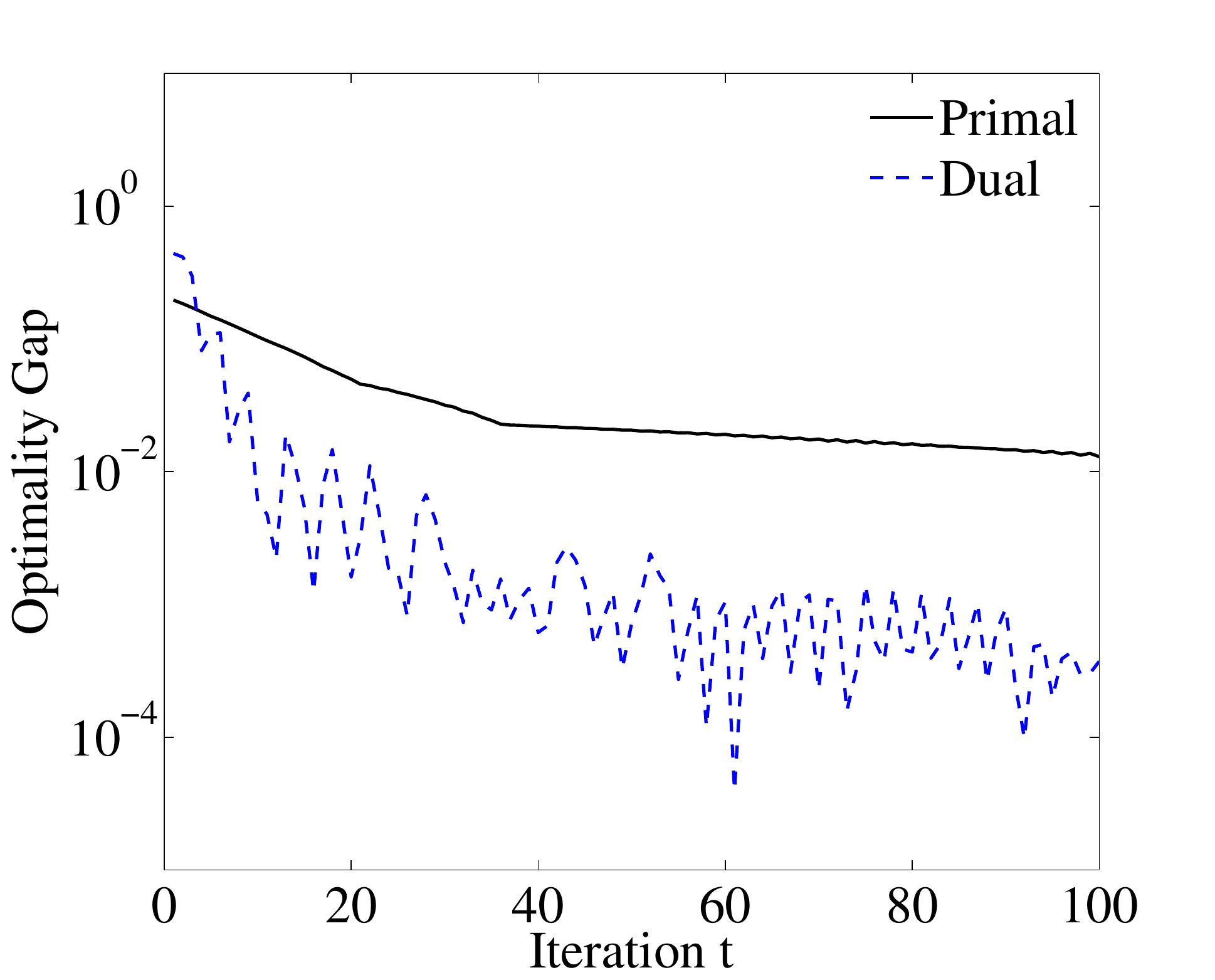}}
\subfloat[]{\includegraphics[width=.51\columnwidth]{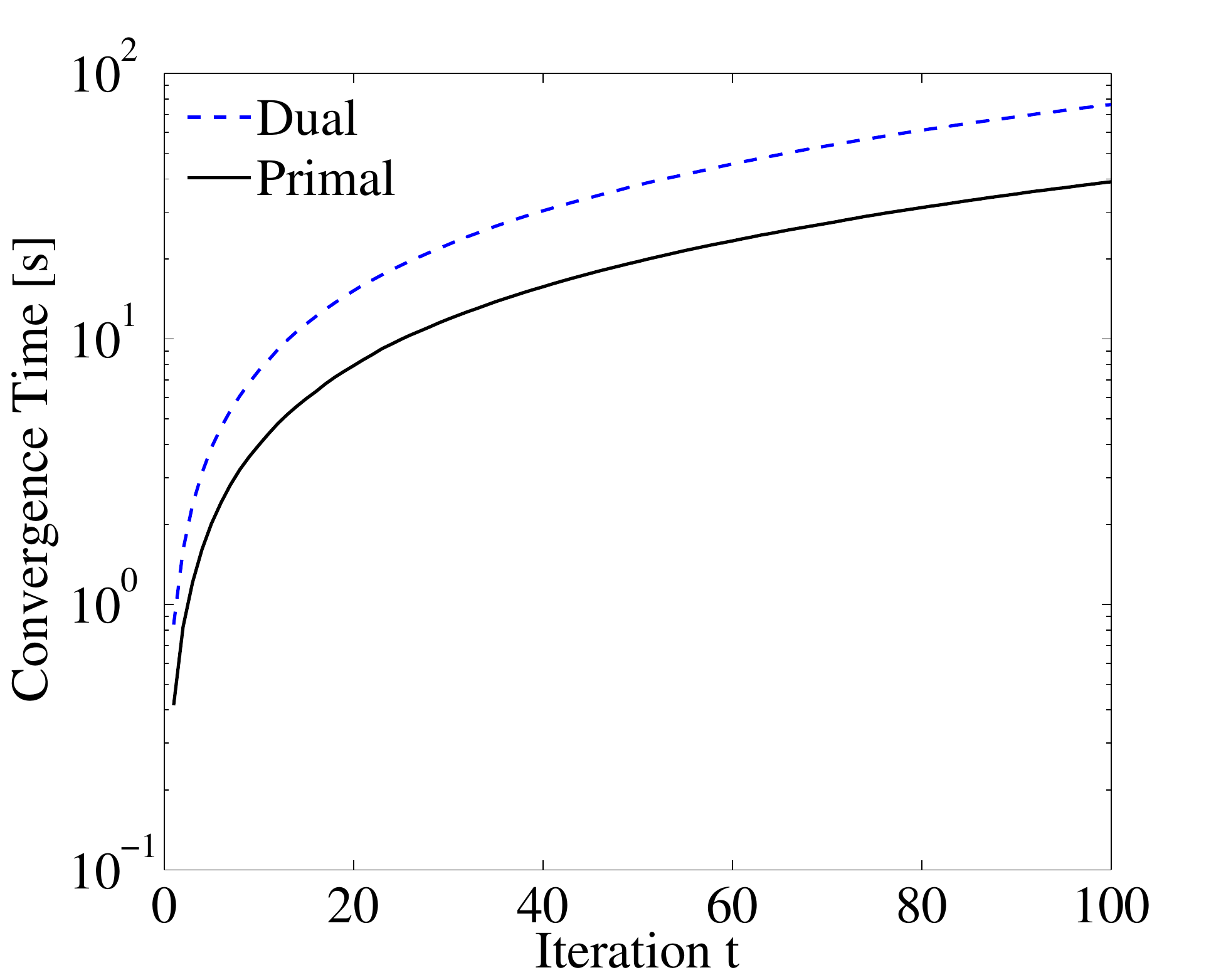}}
\subfloat[]{\includegraphics[width=.47\columnwidth]{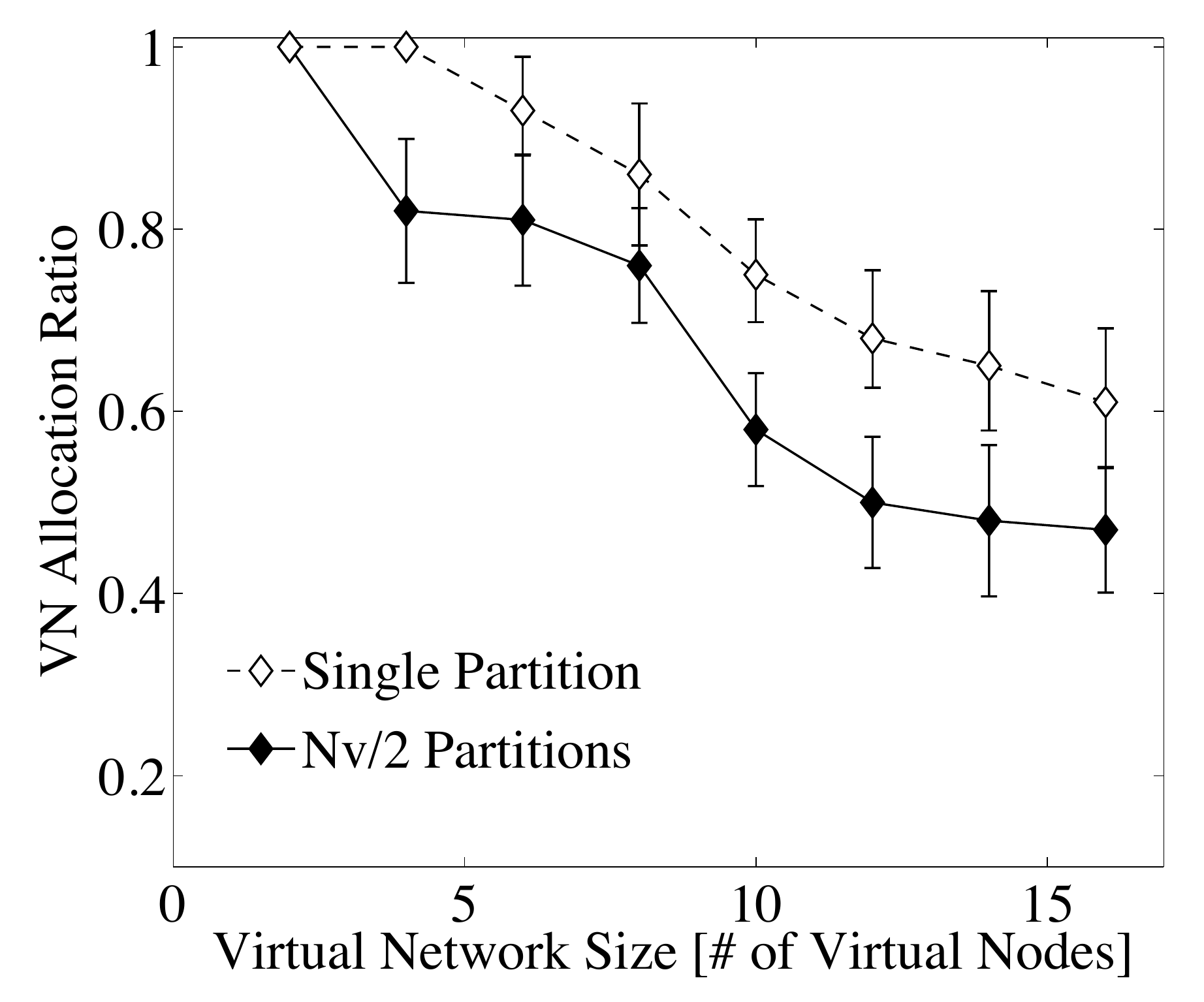}}
\subfloat[]{\includegraphics[width=.48\columnwidth]{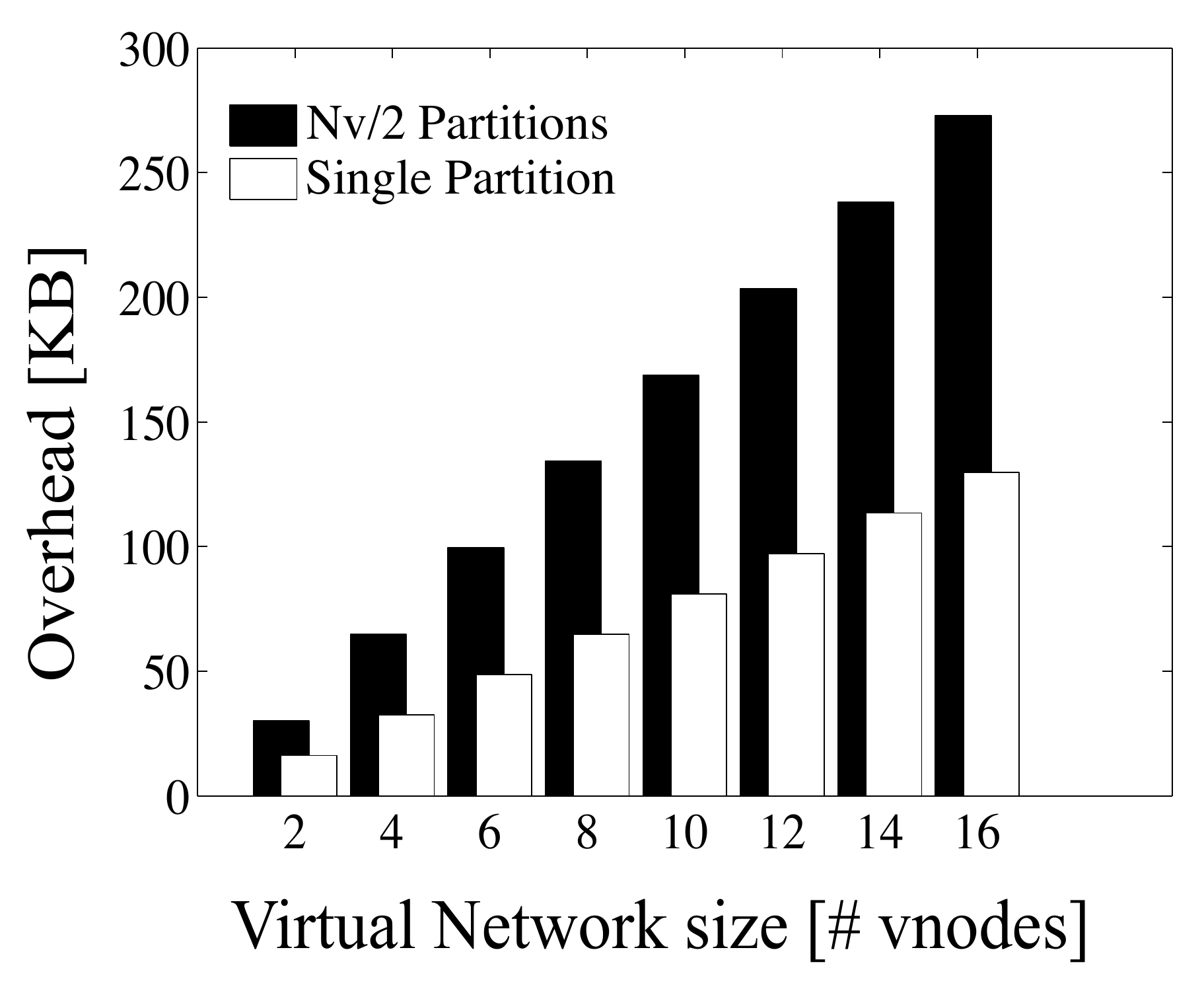}}
\vspace{-2mm}
\caption{(a-b) {\bf Simulations.} Using a diminishing step size rule $\alpha_t = 0.5/t$ to complete the first $100$ iterations, a node embedding solved by dual decomposition leads to a smaller duality gap (a), at the expense of a longer convergence time. (c-d) {\bf Prototype evaluation}. In a primal decomposition, partitioning a VN resulted in a lower VN request allocation ratio, leading to lower cloud revenue (c), and to higher signaling overhead (d). }
\label{figura}
\vspace{-6mm}
\end{figure*}

%
We solve problem~(\ref{dual})  using the projected subgradient method~\cite{boyd-book}.
To find a subgradient of $q$ at $\lambda$,  we let $u^{\star}$ and $v^{\star}$ be the optimal solutions of the subproblems:
\begin{subequations}
\begin{eqnarray} 
u^{\star} = \max_u & (F^T\lambda +c)^T u \\
{\rm  subject \; \rm to} & Au \leq b
\label{subproblem1}
\end{eqnarray}
\end{subequations}   
\vspace{-2mm}
and
\vspace{-2mm}
\begin{subequations}\label{primal-general}
\begin{eqnarray} 
v^{\star} = \max_v & (\tilde{F}^T\lambda +\tilde{c})^T v \\
{\rm  subject \; \rm to} & \tilde{A}v \leq \tilde{b}
\label{subproblem2}
\end{eqnarray}
\end{subequations}
respectively.  Then,  the infrastructure provider processes in charge of solving the subproblems send their optimal values to the master problem, so that the subgradient of the dual function can be computed as:
\begin{equation}
g = F u^{\star} + \tilde{F} v^{\star} - h. 
\end{equation}
%
The subgradient method is run until a termination condition is satisfied (Procedure~\ref{alg:dual}); the operator $(\cdot )_{+}$ denotes the non-negative part of a vector, $i.e.$, the projection onto the non-negative orthant.
\floatname{algorithm}{Procedure}
\begin{algorithm}[t]                   
\caption{\small{Distributed Embedding by Dual Decomposition}}
\begin{algorithmic}[1]              
\STATE {Given $\lambda_{t}$ at iteration $t$, solve the subproblems to obtain the optimal values $u^{\star}$ and $v^{\star}$ for each VN partition}
\STATE {Send/Receive optimal node embedding $u^{\star}$ and $v^{\star}$ }
\STATE{Master computes the subgradient $g = F u^{\star} + \tilde{F} v^{\star} - h$ }
\STATE{Master updates the prices $\lambda_{t+1} = (\lambda_t - \alpha_t g)_{+}$}
\end{algorithmic}
\label{alg:dual}
\end{algorithm} 

At each step, the master problem sets the prices for the virtual nodes to embed. 
The subgradient $g$ 
%
in this case represents the margin of the original shared coupling constraint. If the subgradient associated with the capacity of physical node $i$ is positive ($g_i > 0$), then it is possible for the two subsystems to use more physical capacity of the physical node $i$. 
The master algorithm adjusts the price vector so that the price of each overused physical node is increased, and the price of each underutilized physical node is decreased, but never negative.

\section{Embedding Policies Evaluation}\label{sec:eval}
\vspace{-1mm}
Using both simulations and our downloadable Linux-based embedding testbed, detailed in Chapter $6$ of~\cite{myPhDThesisTR}, we evaluate few representative decomposition policies of our architecture.
{\bf Simulation results.} Our simulations use a CPLEX solver to analyze the tradeoff between optimality and the speed of convergence of the primal and dual decompositions solved by the iterative methods described in Procedures~\ref{alg:primal} and \ref{alg:dual}.  We embed a typical VN request of $50$ virtual nodes onto a physical network overlay of $10$ physical (hosting) nodes. These values are picked as average of an $8$-year dataset of real VN embedding requests to the Emulab VN testbed~\cite{emulab}. The dataset is described in Chapter $4$ of \cite{myPhDThesisTR}.
Since we can always embed a VN leaving no residual capacity on the hosting nodes, the {\it Slater condition}~\cite{boyd-book} is satisfied for Problems~(\ref{masterPrimal}) and (\ref{dual}). This means that there is no duality gap, but it is not desirable to wait for the optimal node embedding when the improvements relative to the previous iterations are small. Hence, using a diminishing step size rule $\alpha=0.5/t$, where $t$ is the iteration step, we stopped our simulations after $t = 100$ (Figures~\ref{figura}$a$ and $b$). We note that the solutions found using a dual decomposition policy reduces faster its duality gap, at the expense of a longer convergence time.  

\noindent
{\bf Prototype evaluation.} We also study the impact of the VN partitioning policy in a primal decomposition, using our VN embedding testbed. Our system is a host running an Ubuntu distribution of Linux (v.$12.04$).  A physical network overlay is emulated via TCP connections on the host loopback interface. 
Each emulated virtual node is a user-level process that has its own virtual Ethernet interface(s), created and installed with {\tt ip link add/set}, and attached to an Open vSwitch~\cite{openvswitch} running in kernel mode to switch packets across virtual interfaces. 
A virtual link is a virtual Ethernet (or {\tt veth}) pair, that acts like a wire connecting two virtual interfaces, or virtual switch ports. Packets sent through one interface are delivered to the other, and each interface appears as a fully functional Ethernet port to all system and application software. The data rate of each virtual link is enforced by Linux Traffic Control (tc), which has a number of packet schedulers to shape traffic to a configured rate. 
During our experiments, the Ubuntu image was hosted on a VirtualBox instance within a $2.5$ GHz Intel Core i5 processor, with 4GB of DDR3 memory. We use the Google Protocol Buffer~\cite{GoogleProtocolBuffer} as a specification language to define constrained VN requests. We analyze the impact that a partitioning policy has on both the allocation ratio, $i.e.$, the ratio between VN allocated on the physical network and requested, and the overhead, $i.e.$ the signaling required to run a distributed VN embedding algorithm using a primal decomposition. 

We attempt to embed $100$ VNs, with a random topology (virtual link exists with probability $p =0.5$), on a fully connected physical network of $5$ physical nodes.
We run the emulation without partitioning the VN, and with a partitioning policy of $N_v/2$, where $N_v$ is the number of requested virtual nodes. In contrast with recent VN embedding solutions~\cite{Houidi2011}, that propose VN partitioning as a necessary step for provisioning a VN, we found that, under the described settings, partitioning a VN not only increases the signaling overhead (Figure~\ref{figura}$d$), but decreases the VN allocation ratio, and therefore the provider revenue (Figure~\ref{figura}$c$). Intuitively, this is because connecting  embedded partitions unbalance the physical network load, reducing the space of feasible solutions for future VN requests.  By exploring the parameter space, we also found that partitioning may lead to higher VN allocation ratios if the physical network is linear (results not shown).   Note that even when a VN is not partitioned, the distributed iterative method used for either primal or dual decompositions implies a message passing between the master and the dual subproblems. Partitioning a VN means running such iterative method multiple ($N_v/2$) times, but on problems with smaller input size.

\vspace{-2mm}
\section{Conclusions}
\vspace{-1mm}

The virtual network (VN) embedding protocol is one of the crucial, yet not standardized protocols that cloud providers need to run in support of wide-area virtualization-based services.
In this paper, we proposed an architecture that leverages decomposition theory to provide insights into a systematic design of distributed VN embedding solutions. 
We model the three interacting mechanisms of the virtual network embedding problem ---physical resource discovery, virtual network mapping, and allocation--- with separate subproblems, interfaced with the optimization variables that coordinate each subproblem. 
Using both simulations and our VN embedding prototype, we showed how our architecture can be used to analyze key VN embedding protocol design tradeoffs. Using our CPLEX-based simulator, we found how some decompositions may lead to quicker reduction of the optimality gap, at the expense of a slower speed of the distributed iterative method. Furthermore, we used our Linux-based VN embedding testbed to assess the impact of different VN partitioning policies for primal decompositions, under two performance metrics: the VN allocation ratio, and the signaling overhead. Our use case study demonstrates how our architecture can be used to design customized distributed VN embedding solutions by policy instantiation.

\setstretch{0.93}
\bibliographystyle{abbrv}
\bibliography{all}

\begin{thebibliography}{10}

\bibitem{GENI}
{The GENI initiative} \url{http://www.geni.net}.

\bibitem{sword-journal}
J.~Albrecht, D.~Oppenheimer, A.~Vahdat, and D.~A. Patterson.
\newblock {Design and Implementation Trade-offs for Wide-area Resource
  Discovery}.
\newblock {\em ACM Transaction Internet Technologies}, 8(4):1--44, 2008.

\bibitem{bellagio}
AuYoung, Chun, Snoeren, and Vahdat.
\newblock {Resource Allocation in Federated Distributed Computing
  Infrastructures}.
\newblock {\em In Proc. of Workshop on OS and Arch. Support for the On demand
  IT Infrastr.}, October 2004.

\bibitem{azer-ocx}
B.~Azer and K.~Orran.
\newblock {Towards an Open Cloud Marketplace: Vision and First Steps}.
\newblock In {\em IEEE Internet Computing Magazine}, Jan 2014.

\bibitem{boyd-book}
S.~Boyd and L.~Vandenberghe.
\newblock {\em Convex Optimization}.
\newblock http://www.stanford.edu/people/boyd/cvxbook.html, 2004.

\bibitem{ChowdhuryTON}
M.~Chowdhury, M.~R. Rahman, and R.~Boutaba.
\newblock {ViNEYard: Virtual Network Embedding Algorithms with Coordinated Node
  and Link Mapping}.
\newblock {\em IEEE/ACM Trans. Netw.}, 20(1):206--219, Feb. 2012.

\bibitem{Polyvine}
M.~Chowdhury, F.~Samuel, and R.~Boutaba.
\newblock {PolyViNE: Policy-Based Virtual Network Embedding Across Multiple
  Domains}.
\newblock SIGCOMM VISA '10 Workshop, pages 49--56, New York, NY, USA, 2010.
  ACM.

\bibitem{mappingNPhard}
B.~Chun and A.~Vahdat.
\newblock {Workload and Failure Characterization on a Large-scale Federated
  Testbed.}
\newblock Technical report, IRB-TR-03-040, Intel Research Berkeley, 2003.

\bibitem{myPhDThesisTR}
F.~Esposito.
\newblock {\em {A Policy-based Architecture for Virtual Network Embedding}}.
\newblock PhD thesis, Computer Science Department, Boston University. Technical
  Report CS-TR-2013-012, Sept. 2013.

\bibitem{CAD}
F.~Esposito, D.~{Di Paola}, and I.~Matta.
\newblock {A General Distributed Approach to Slice Embedding with Guarantees}.
\newblock In {\em {Proc. of the IFIP Networking}}, Brooklyn, NY, USA, 2013.

\bibitem{SHARP}
Y.~Fu, J.~Chase, B.~Chun, S.~Schwab, and A.~Vahdat.
\newblock {SHARP: An Architecture for Secure Resource Peering}.
\newblock {\em SIGOPS Operating System Review}, 37(5):133--148, 2003.

\bibitem{GoogleProtocolBuffer}
{Google Protocol Buffer}.
\newblock {Developer Guide} \url{http://code.google.com/apis/protocolbuffers/},
  2010.

\bibitem{Houidi2011}
I.~Houidi, W.~Louati, W.~Ben~Ameur, and D.~Zeghlache.
\newblock Virtual {N}etwork {P}rovisioning across {M}ultiple {S}ubstrate
  {N}etworks.
\newblock {\em Computer Networks}, 55(4):1011--1023, Mar. 2011.

\bibitem{Houidi-distributedVNM}
I.~Houidi, W.~Louati, and D.~Zeghlache.
\newblock {A Distributed Virtual Network Mapping Algorithm}.
\newblock In {\em IEEE International Conference on Communications (ICC)}, pages
  5634 --5640, May 2008.

\bibitem{isomorphism-visa09}
J.~Lischka and H.~Karl.
\newblock {A Virtual Network Mapping Algorithm based on Subgraph Isomorphism
  Detection}.
\newblock {\em VISA, ACM SIGCOMM Workshop}, 17 August 2009.

\bibitem{openvswitch}
{Open Virtual Switch}.
\newblock \url{http://www.openvswitch.org}.

\bibitem{sword}
D.~Oppenheimer, J.~Albrecht, D.~Patterson, and A.~Vahdat.
\newblock {Design and Implementation Tradeoffs for Wide-Area Resource
  Discovery}.
\newblock {\em HPDC, High Performance Distributed Computing}, 2005.

\bibitem{emulab}
B.~e.~a. White.
\newblock {An Integrated Experimental Environment for Distributed Systems and
  Networks}.
\newblock {\em SIGOPS Oper. Syst. Rev.}, 36(SI):255--270, 2002.

\bibitem{pathsplitting}
M.~Yu, Y.~Yi, J.~Rexford, and M.~Chiang.
\newblock {Rethinking Virtual Network Embedding: Substrate Support for Path
  Splitting and Migration}.
\newblock {\em SIGCOMM Comput. Commun. Rev.}, 38(2):17--29, 2008.

\bibitem{V-Mart}
F.~Zaheer, J.~Xiao, and R.~Boutaba.
\newblock {Multi-provider Service Negotiation and Contracting in Network
  Virtualization}.
\newblock In {\em IEEE Network Oper. and Management Symposium (NOMS)}, pages
  471 --478, April 2010.

\bibitem{cabernet}
Y.~Zhu, R.~Zhang-Shen, S.~Rangarajan, and J.~Rexford.
\newblock {Cabernet: Connectivity Architecture for Better Network Services}.
\newblock CoNEXT, pages 64:1--64:6. ACM, 2008.

\end{thebibliography}

\end{document}